\documentclass[aps, prd, preprint, nofootinbib]{revtex4}

\usepackage{amssymb, amsmath} 
\usepackage{stmaryrd}

\usepackage{bbm} 

\usepackage[english]{babel}
\usepackage{slashed}
\usepackage{cancel}

\usepackage{slashbox}

\usepackage{graphicx}
\usepackage{color}

\usepackage{datetime}
\settimeformat{ampmtime}

\usepackage{float}

\usepackage[bookmarks,breaklinks,plainpages=false,pdfpagelabels]{hyperref} 
	\hypersetup{linkcolor=red,citecolor=blue,filecolor=dullmagenta,urlcolor=darkblue} 

\usepackage{setspace}
\definecolor{grey}{rgb}{0.4,0.4,0.4}
\definecolor{lightgrey}{rgb}{0.6,0.6,0.6}
\definecolor{dullmagenta}{rgb}{0.4,0,0.4}
\definecolor{darkblue}{rgb}{0,0,0.4}
\definecolor{orange}{rgb}{1,0.5,0}
\definecolor{lightbrown}{rgb}{0.75,0.5,0.25}
\definecolor{tan}{cmyk}{0.14,0.42,0.56,0}
\definecolor{djunglegreen}{cmyk}{0.99,0,0.52,0}
\definecolor{lightgreen}{rgb}{0,1,0}
\definecolor{olivegreen}{cmyk}{0.64,0,0.95,0.40}
\definecolor{midgreen}{rgb}{0.0,0.675,0.0}

\newcommand{\vs}{\vspace}

\renewcommand{\.}{\hspace{0.5mm}}

\newcommand{\ra}{\ensuremath{\rightarrow}}

\newcommand{\irm}{\ensuremath{\mathrm{i}}}

\newcommand{\Hcal}{\ensuremath{\mathcal{H}}}

\renewcommand{\d}{\ensuremath{\mathrm{d}}}

\newcommand{\defas}{\mathrel{\mathop :}=} 
\newcommand{\hph}[1]{\hphantom{#1\;\,}}

\newcommand{\eg}{e.g.}
\newcommand{\ie}{i.e.}

\newcommand{\rhs}{r.h.s.}

\newcommand{\cf}{cf.}

\makeatletter
 \def\ifundefined#1{\expandafter\ifx\csname#1\endcsname\relax}
 \ifundefined{default@color}
  \let\default@color=\current@color
 \fi
\makeatother

\newcommand{\beq}{\begin{equation}}
\newcommand{\eeq}{\end{equation}}
\newcommand{\bea}{\begin{eqnarray}}
\newcommand{\beas}{\begin{eqnarray*}}

\newcommand{\eea}{\end{eqnarray}}
\newcommand{\eeas}{\end{eqnarray*}}

\begin{document}

\title{Thoughts on the Vacuum Energy in the Quantum $N$-Portrait}

\author{Florian K{\"u}hnel}
	\email{florian.kuhnel@fysik.su.se}
	\affiliation{The Oskar Klein Centre for Cosmoparticle Physics,
			Department of Physics,
			Stockholm University,
			AlbaNova,
			106\.91 Stockholm,
			Sweden}

\date{\formatdate{\day}{\month}{\year}, \currenttime}

\begin{abstract}
An application of the quantum $N$-portrait to the Universe is discussed, wherein the space-time geometry is understood as a Bose-Einstein condensate of $N$ soft gravitons. If near or at the critical point of a quantum phase transition, indications are found that the vacuum energy is partly suppressed by $1 / N$, as being due to quanta not in the condensate state. Time evolution decreases this suppression, which might have implications for cosmic expansion.
\end{abstract}


\maketitle

\noindent{{\color{white}.}\\[-9.5mm]\it Introduction\,---\!}
The theory of General Relativity provides a cornerstone of classical physics. It is remarkably simple and relates gravitational effects to the curvature of space time. On that level it is very well understood and tested on a broad range of length scales. However, as soon as one starts to combine it with quantum mechanics\./\.quantum field theory its understanding was rather poor, and even on a semi-classical level lead to various mysteries, in particular in the context of black holes, like the information paradox, the negativ heat capacity, the Bekenstein entropy etc. Only recently a consistent picture of geometry on the quantum level has been developed which resolves the mentioned paradoxes \cite{Dvali:2012wq, Dvali:2, Dvali:2013eja, Dvali, Dvali2, Dvali3, Dvali4, Dvali5, Dvali6, Dvali7, Dvali8, Dvali9, Dvali10}. The key idea is that space-time geometry is made out of a self-sustained Bose-Einstein condensate of gravitons. Mainly applied to black holes, a variety of essential black-hole features have been successfully described by estimates in this language, such as the ones mentioned above (\cf~also Refs.~\cite{Dvali:2014gua, Casadio:2015bna, Flassig:2013, N-Portrait-related, N-Portrait-related2, Brustein, Brustein2, Brustein3, Casadio, Casadio2, Casadio:2015xva,  Kuhnel-Sundborg-1, Kuhnel:2015qaa}). Recently, the quantum $N$-portrait has been applied to inflation \cite{Dvali:2013eja, Casadio:2015xva, Kuhnel:2015yka}, and it has been shown that new quantum effects are expected (affecting, for instance, the total number of e-foldings and the tensor-to-scalar ratio). It is important to note that those features are not captured by any semi-classical treatment.

More formal investigations are based on scalar toy models which resemble remarkable well certain features of black holes that are expected from the treatment of space-time geometry as a graviton Bose-Einstein condensate (\cf~Refs.~\cite{Dvali:2, Flassig:2013, N-Portrait-related, N-Portrait-related2, Kuhnel-Sundborg-2}). Now, precisely those models are also used in the context of analogue gravity (\cf~Refs.~\cite{Unruh:1981bi, Visser:1993, Visser:1997ux, BLV:2001, Girelli:2008gc, FRW}, and \cite{Barcelo-Living-Review} for an extended review), wherein geometry emerges in an effective treatment of fluctuations in the condensate. Various types of emergent geometries have been studied, especially black holes in various facets \cite{Unruh:1981bi, Visser:1993, Visser:1997ux}, but also Friedmann-Lema{\^i}tre-Robertson-Walker geometries \cite{FRW}, just to mention two (\cf~Ref.~\cite{Barcelo-Living-Review} and references therein). In most of the works in the field of analogue gravity, the analogy is purely on the kinematical level in the sense that fluctuations (for instance of the velocity potential) evolve in an effective geometry sourced by the system's background structure. Although, so far, there has not been developed a true dynamical analog resembling fully-relativistic Einstein gravity, its non-relativistic limit and varieties thereof have been established on the dynamical level. For instance, in \cite{Girelli:2008gc} it was shown that a modified Poisson equation is encoded in a scalar field model with quartic self-interactions, and this is one model which was used as a major toy model for studying features of the quantum $N$-portrait (\cf~Refs.~\cite{Dvali:2, Flassig:2013}).

The purpose of this letter is now to combine those two approaches, understanding space-time geometry as a graviton Bose-Einstein condensate and analogue gravity. We will first briefly review the toy model originally used by Dvali and Gomez \cite{Dvali:2}, and review some of its features. Next, we will relate it to the results of \cite{Girelli:2008gc}, and then discuss its implications on the vacuum energy, when the framework is applied to the observable Universe.\\[-4mm]

\noindent{\it Set-Up\,---\!}
Following \cite{Dvali:2}, our starting point shall be the Hamiltonian
\vs{-1mm}
\begin{align}
	H
		&=
								-
								\int\limits_{V}\! \d^{3} x
								\Big[
									\hslash\.L
									\Psi^{\dagger}\.\nabla^{2}\Psi
									+
									g\.\Psi^{\dagger}\.\Psi^{\dagger}\.\Psi\.\Psi
								\Big]
								,
								\label{eq:H}
								\\[-8.3mm]
		&
								\notag
\end{align}
where $\Psi$ is a field operator describing the order parameter of a Bose gas, $L$ is a parameter of length-dimensionality, and $g$ is the coupling constant of the two-body interaction. We assume that the system is in a finite box of size $R$ and volume $V = ( 2 \pi R )^{3}$, with periodic boundary conditions $\Psi( 0 ) = \Psi( 2 \pi R )$. The total particle number shall be $N$. In Ref.~\cite{Dvali:2} it has been argued that if
\vs{-3.5mm}
\begin{align}
	\alpha
		&=
								\frac{ 1 }{ N }
								\; ,
								\label{eq:alpha=1/N}
\end{align}
where $\alpha \defas 2\.g\.R^{2} / ( \hslash\.V L )$, the system is at the critical point of a quantum phase transition.
\newpage

\noindent Furthermore, the first Bogoliubov energy level, $\epsilon_{1}$, fulfills
\begin{align}
	\tilde{\epsilon}_{1}
		&=
								\frac{ 1 }{ N }
								\; ,
								\label{eq:epsilontilde1=1/N}
\end{align}
with $\tilde{\epsilon}_{1} \defas L_{\text{P}}\.\epsilon_{1} / \hslash$ and $L_{P} \defas L / \sqrt{N\,}$ (expressing {\it maximal packing}, \cf~Ref.~\cite{Dvali:2}). Evidently, the energy gap \eqref{eq:epsilontilde1=1/N} goes to zero in the limit of $N \ra \infty$ (\cf~\cite{Flassig:2013} for a numerical computation of the energy spectrum). It has been argued \cite{Dvali:2} that this collapse of the energy levels might be the explanation of the black-hole information paradox. Also, Hawking radiation is simply understood as quantum depletion of the condensate. We will come back to the last point at the end of this section.

The mentioned collapse of the energy spectrum for the chosen value of the self-coupling 
has been numerically investigated in $1 + 1$ dimensions in Ref.~\cite{Flassig:2013}. Therefore, a $U( 1 )$-breaking term has been added in order to localize the solitonic solutions of one phase of the system.

Now, one can show that the model \eqref{eq:H} encodes a modified Poisson equation for the (analogue gravity) potential in which the condensate excitations evolve (see below), thus establishing a non-relativistic dynamical gravitational analogy. Again, this derivation is only possible with breaking of the $U( 1 )$ symmetry. We follow Ref.~\cite{Girelli:2008gc} and add
\begin{align}
	H_{\text{SB}}
		&=
								-\;
								\frac{ \lambda }{ 2 }
								\int\limits_{V}\! \d^{3} x
								\left[
									\Psi\.\Psi
									+
									\Psi^{\dagger}\.\Psi^{\dagger}
								\right]
								,
								\label{eq:U1breaking}
\end{align}
to the Hamiltonian \eqref{eq:H}, which gives mass to the excitations. The term $H_{\text{SB}}$ is the simplest one which leaves the expectation value of the number operator constant. 
Collecting all terms, the grand-canonical Hamiltonian reads
\begin{align}
	{\Hcal}
		&=
								\int\limits_{V}\! \d^{3} x
								\bigg[
									-
									\hslash\.L\.
									\Psi^{\dagger}\.\nabla^{2}\Psi
									+
									g\.\Psi^{\dagger}\.
									\Psi^{\dagger}\.\Psi\.\Psi
									-
									\mu \Psi^{\dagger}\Psi
									-
									\frac{ \lambda }{ 2 }
									\left(
										\Psi\Psi
										+
										\Psi^{\dagger}\.\Psi^{\dagger}
									\right)\!
								\bigg]
								,
								\label{eq:Hfull}
\end{align}
with $\mu$ being the chemical potential which has the same dimension as $\lambda$.

The first thing one might note is that the Heisenberg equation of motion of this model possesses a static and homogeneous solution,
\begin{align}
	\Psi_{0}
		&=
								\sqrt{\frac{ \mu + \lambda }{ 2\.g }\.}
								\; .
								\label{eq:Psi-0}
\end{align}

Next, we decompose $\Psi$ into the classical background $\Psi_{0}$ and a quantum-fluctuation part $\delta \Psi$ (obeying canonical commutation relations),
\begin{align}
	\Psi
		&\equiv
								\Psi_{0}
								+
								\delta \Psi
								\; .
								\label{eq:decomposition}
\end{align}
Since in this work we shall mainly be interested in vacuum contributions, we evaluate the Heisenberg equation of motion for $\Psi$ in a state with no excitations, $| 0 \rangle$. This leads to a generalized Bogoliubov--De Gennes equation for the condensate wave function,
\vs{-1mm}
\begin{align}
\begin{split}
	\irm \hslash \frac{\partial \Psi_{0}}{\partial t}
		&\simeq
								-
								\hslash^{2}\.L\.\nabla^{2}\Psi_{0}
								-
								\mu\.\Psi_{0}
								-
								\lambda \Psi_{0}^{*}
								+
								2\.g\.\big| \Psi_{0} \big|^{2}\.\Psi_{0}
								\\[1mm]
		&\hph{\simeq}
								+
								4\.g\.
								\big\langle 0 \big|
									\delta \Psi^{\dagger}\.\delta \Psi
								\big| 0 \big\rangle\.
								\Psi_{0}
								+
								2\.g\.
								\big\langle 0 \big|
									\delta \Psi^{2}
								\big| 0 \big\rangle\.
								\Psi_{0}^{*}
								\; .
								\label{eq:BdG-eq}
\end{split}
\end{align}

Before discussing further the background part, we now turn to the equation of motion for the fluctuation $\delta \Psi$ in the static case. Parametrizing deviations from the homogeneous background via
\vs{-2mm}
\begin{align}
	\Psi_{0}( x )
		&=
								\sqrt{\frac{ \mu + \lambda }{ 2\.g }\.}\.
								\Bigg[
									1
									+
									c\,
									\Phi_{\text{g}}( x )
								\Bigg]
								\; ,
								\label{eq:non-homogeneous-background}
\end{align}
for some constant $c( \mu, \lambda )$, leads to
\begin{subequations}
\begin{align}
	0
		&\simeq
								-
								\hslash^{2}\.L\.\nabla^{2}\delta \Psi
								+
								\big[
									( \mu + 2 \lambda)
									+
									4 (\mu + \lambda)\.c^{2}\,
									\Phi_{\text{g}}
								\big]
								\delta \Psi
								+
								\big[
									\mu
									+
									2 (\mu + \lambda)\.c^{2}\,
									\Phi_{\text{g}}
								\big]
								\delta \Psi^{\dagger}
								\; ,
								\label{eq:deltaPsi-EOM}
								\\[2mm]
	0
		&\simeq
								-
								\hslash^{2}\.L\.\nabla^{2}\delta \Psi^{\dagger}
								+
								\big[
									( \mu + 2 \lambda)
									+
									4 (\mu + \lambda)\.c^{2}\,
									\Phi_{\text{g}}
								\big]
								\delta \Psi^{\dagger}
								+
								\big[
									\mu
									+
									2 (\mu + \lambda)\.c^{2}\,
									\Phi_{\text{g}}
								\big]
								\delta \Psi
								\; .
								\label{eq:deltaPsi-EOM-dagger}
\end{align}
\end{subequations}
Combining those two equations to diagonalization for $\delta \Psi$, one finds \cite{Girelli:2008gc} that, at low momenta and for small $\Phi_{\text{g}}$, the excitations are governed by the Hamiltonian
\vs{-1mm}
\begin{align}
	H_{1}
		&\simeq
								-
								\frac{ \hslash^{2} }{ 2\.M }\.\nabla^{2}
								+
								M c_{s}^{2}
								+
								M\.\Phi_{\text{g}}
								\; ,
								\label{eq:H-qp}
\end{align}
where 
$c_{s} = c_{s}( \mu, \lambda )$ and $M = M( \mu, \lambda )$. The Hamiltonian \eqref{eq:H-qp} describes a non-relativistic particle of mass $M$, in a potential $\Phi_{\text{g}}$, which satisfies a modified (short-range) Poisson equation (\cf~Ref.~\cite{Girelli:2008gc})
\begin{align}
	\left[
		\nabla^{2}
		-
		\frac{2\.( \mu + \lambda )}{\hslash^{2}\.L}
	\right]\!
	\Phi_{\text{g}}
		&=
								g\,
								\frac{4\.
								(
									\mu
									+
									4\.\lambda
								)
								(
									\mu
									+
									2\.\lambda
								)}
								{\hslash^{2}\.\lambda}\,
								\bigg(
									\big\langle 0 \big|
										\delta \Psi^{\dagger}\.\delta \Psi
									\big| 0 \big\rangle
									+
									\frac{ 1 }{ 2 }\.
									\big\langle 0 \big|
										\delta \Psi^{2}
									\big| 0 \big\rangle
								\bigg)
								\, .
								\label{eq:poisson}
\end{align}
Note that the term on the \rhs~is a pure vacuum contribution. In general, when one considers the equation of motion on different non-vacuum states, and also interactions with other fields, $\Phi_{i}$, than $\Psi$ (which represents gravitons here), the resulting Poisson-like equation has the structure
\vs{-2mm}
\begin{align}
	\left[
		\nabla^{2}
		-
		\frac{ 1 }{ \ell^{2} }
	\right]\!
	\Phi_{\text{g}}
		&\simeq
								4 \pi\.G_{\text{N}}\,
								\rho_{\text{m}}( \Psi, \Phi_{i} )
								+
								\Lambda( \Psi, \Phi_{i} )
								\; ,
								\label{eq:general-poisson}
\end{align}
for some length scale $\ell$ and for a suitably defined analog Newton constant $G_{\text{N}}$. Both are built from background quantities, and are hence{\,---\,}like the analogue Planck mass $M_{\text{P}}${\,---\,}emergent. The concrete structure of $\rho_{\text{m}}( \Psi, \Phi_{i} )$ depends on the specific model under consideration. In general it contains expectation values of the fluctuations of the various fields, evaluated on a particular state, $| \zeta \rangle$. For instance, in the case of the Hamiltonian \eqref{eq:Hfull} one has (\cf~Ref.~\cite{Girelli:2008gc})
\begin{align}
	\rho_{\text{m}}( \Psi )
		&\sim
								2\.\langle \zeta | \delta \Psi^{\dagger}\.\delta \Psi | \zeta \rangle
								+
								\langle \zeta | \delta \Psi^{2} | \zeta \rangle
								-
								( \text{vacuum contribution} )
								\; .
								\label{eq:rho-m}
\end{align}

Let us now built the bridge to the graviton Bose-Einstein-condensate picture of Ref.~\cite{Dvali:2012wq}. This has, as mentioned at the beginning of this section, in a basic approach the same Hamiltonian as that used above in the discussion on analogue gravity. The key feature of the graviton case is, however, its particular value of the self-coupling $g \sim 1 / N$, which is such that the system is (and always stays) at the critical point of a quantum phase transition. Now, as can be read off from Eq.~\eqref{eq:poisson}, the cosmological constant $\Lambda$ is proportional to $g$, and hence we find
\vs{-2mm}
\begin{align}
	\Lambda
		&\sim
								\frac{ 1 }{ N }
								\; .
								\label{eq:CLambdasim1/N}
\end{align}
This shows that the contribution of the vacuum-energy density is strongly suppressed if $N$ is large. If the set-up under consideration could be applied to the observable Universe, one might face a suppression of $\Lambda$ by the number of gravitons making up this background geometry, being of the order $1 / N \approx 10^{-120}$.

It is interesting to note that the $1 / N$-behavior of $\Lambda$ coincides with that of the Bogoliubov energy levels $\tilde{\epsilon}_{i}$ [\cf~Eq.~\eqref{eq:epsilontilde1=1/N}]. This is, however, clear when one considers the fact that the vacuum contributions are due to quanta {\it outside} of the condensate, \ie~due to depletion [\cf~Eq.~\eqref{eq:BdG-eq}]. Gravitons that escape the condensate will decrease the mentioned suppression, leading to an even stronger acceleration, and this will have consequences on the time evolution of cosmic expansion. Furthermore, the more quanta deplete from the condensate, the more quantum the system becomes, till it eventually does not posses a (semi-)classical description anymore. In this spirit, in Ref.~\cite{Dvali:2013eja} a bound on the total number of e-foldings has been derived.\\[-3mm]

\noindent{\it Summary \& Outlook\,---\!}
In this letter we have established a bridge between the so-called quantum $N$-portrait in which geometry is described as a Bose-Einstein condensate of gravitons, and analogue-gravity systems where the notion of geometry and gravity comes about as an effective\./\.emergent concept in which certain fluctuations evolve. This is possible, since the toy models for the former serve as important fields of application for the latter. For instance, the model studied encodes a modified Poisson equation, thus establishing a non-relativistic dynamical gravitational analogy. Despite of being of short-range nature, it resembles remarkable well the structure one would encounter in the non-relativistic limit of General Relativity, including the presence of a vacuum-energy term. The key observation is the $N$-dependence of this analogue cosmological-constant term, which, provided the system is at the critical point of a quantum phase transition, scales as $1 / N$. If applied to the observable Universe, which contains approximately $10^{120}$ gravitons, leads to {\it predict} a suppression corresponding to $\Lambda_{\text{obs}} / M_{\text{Planck}}^{4}$.\footnote{\setstretch{0.85} For a different line of reasoning it was discussed in Ref.~\cite{Binetruy:2012kx} that {\it the most probable} value for the quantum vacuum energy density might be of the order of the critical energy density as observed.}

Of course, the model \eqref{eq:Hfull} represents just a very simply toy model, resembling certain features of graviton condensates. It is a scalar model, without any derivative self-interactions, and furthermore does not contain any other particles. While we leave the first two points entirely for future investigations, we would like to speculate a bit on the latter. We have seen that the suppression for the vacuum contributions is inversely proportional to the number of quanta. This suggests that condensates of different types and\./\.or shapes will face different suppressions; Interactions among their constituents, \eg~gravitons and inflatons, which {\it enhance} their depletion rate (\cf~Ref.~\cite{Dvali:2013eja}), will lead to a {\it decreased} vacuum energy-density suppression.\\[-3mm]

\noindent{\it Acknowledgements\,---\!}
It is a pleasure to thank Gia Dvali, Cornelius Rampf and Bo Sundborg for helpful discussions. The work of FK was supported by the Swedish Research Council (VR) through the Oskar Klein Centre.


\end{document}